\documentclass{article}

\usepackage[preprint,nonatbib]{neurips_2024}
\usepackage{graphicx}
\usepackage{subcaption} 
\usepackage{amsmath} 
\usepackage{cite}

\usepackage{helvet}

\usepackage[numbers]{natbib}

\usepackage[utf8]{inputenc} 
\usepackage[T1]{fontenc}    
\usepackage{hyperref}       
\usepackage{url}            
\usepackage{booktabs}       
\usepackage{amsfonts}       
\usepackage{nicefrac}       
\usepackage{microtype}      
\usepackage{xcolor}         

\title{Have we reached the beginning of the end \\ for review papers?}

\author{%
  Barry Smyth, P\'{a}draig Cunningham \\
  School of Computer Science\\
  University College Dublin\\
\texttt{\{barry.smyth,padraig.cunningham\}@ucd.ie}}

\begin{document}

\maketitle

\begin{abstract}
Review papers have traditionally enjoyed a high status in academic publishing because of the important role they can play in summarising and synthesising a field of research. They can also attract significantly more citations than primary research papers presenting original research, making them attractive to authors. There has been a dramatic increase in the publication of review papers in recent years, both in raw numbers and as a proportion of overall publication output. In this paper we demonstrate this increase across a wide range of fields of study. We quantify the citation dividend associated with review papers, but also demonstrate that it is declining and discuss the reasons for this decline. We further show that, since the arrival of GenAI tools in 2022 there is evidence of widespread use of GenAI in research paper writing, and we present evidence for a stronger AI signal among review papers compared to primary research papers. We suggest that the potential for GenAI to accelerate and even automate the production review papers will have a further significant impact on their status.
  \end{abstract}

\section{Introduction}

In contrast to regular research papers, reporting original research, review papers are designed to synthesize and summarize prior research providing an overview of a field for researchers new to the area, or to offer a new perspective for future research. Because of this aggregation and synthesis of evidence many consider review papers to have a high status. In fact in medical research, systematic reviews are considered to represent the top of the publication hierarchy \cite{evans2003hierarchy}.

While there is evidence that review papers attract more citations than papers reporting original research \cite{mcmahan2021creative,miranda2018overcitation}, this citation dividend comes at the expense of the original research papers covered in the review \cite{mcmahan2021creative}. This promised citation dividend, and the emergence of software tools to support the production of reviews \cite{marshall2015systematic,viechtbauer2010conducting}, has resulted in a significant increase the volume of review papers \cite{miranda2018overcitation,ioannidis2016mass}. This proliferation of review papers has drawn considerable criticism. The main criticism is that new software tools \cite{kohl2018online,oconnor2024large} make review papers easier to write resulting in an over-production of `low-effort' publications of limited value. There is massive redundancy in such work with some research areas being reviewed multiple times \cite{ioannidis2016mass,hoffmann2021nearly}. 

The arrival of GenAI tools will further  impact the ease with which reviews can be produced. 
Bolanos \emph{et al.} suggested that before the end of the decade, AI-enabled research assistants
will be able to generate comprehensive literature reviews \cite{bolanos2024artificial}. Indeed there is ample evidence that this is already happening \cite{qiu2025completing,cao2025automation}. While the quality of these rapid reviews must be open to question, it is clear that the arrival of GenAI dramatically changes  the literature review landscape. It may be that the wider societal problem of \emph{AI Slop} (low effort, low quality, high volume 
AI-generated content) will become a problem for academic publishing as well. 
Yun \emph{et al.} propose that that arrival of LLMs for automating literature reviews opens the prospect of on-demand synposes on any topic of interest~\cite{yun2023appraising}. 
These issues, coupled with the older criticism that review papers pick up citations at the expense of the original research papers \cite{mcmahan2021creative} have damaged the status of reviews. 

For the purpose of the analysis presented here we consider three types of review paper; Systematic Reviews (SR), Meta-Analyses (MA) and Narrative Reviews (NR). These are described in more detail in section \ref{sec:rev_papers}. In the remainder of this article, we present the results of a large-scale (17 million research papers) scientometrics study of review papers across 23 primary fields of study. Our results confirm that there has been a considerable growth in review papers, both in raw numbers and as a proportion of overall research publications, across most fields of study. We demonstrate that the citation dividend for review papers is real -- SRs and MAs perform particularly well -- but declining. The decline is likely linked to the proliferation of reviews, and we provide evidence that this is associated with the recent appearance of several tools to support and accelerate the review process. Looking to the future, we demonstrate increasing use of GenAI in research paper writing, in general, and show evidence of a stronger AI signal in review papers, in particular, which may foreshadow a far greater disruption in the evolution of scientific publishing yet to come.

\section{What are review papers?} \label{sec:rev_papers}

The purpose of review papers is to provide a synthesised overview of the state of research on a particular topic. They can be invaluable for researchers starting out in a new research area because, in addition to this overview, they can highlight gaps and identify promising future research directions \cite{miranda2018overcitation}. Review papers are considered `secondary' sources with the papers covered by the review being the `primary' sources. As such review papers are not expected to report original research.  

Review papers can be subdivided into numerous categories; for the purpose of our analysis we consider three main categories:

\begin{itemize}
    \item Systematic reviews that follow a formal methodology. 
    \item Meta-analysis: a quantitative synthesis of a number of primary studies. 
    \item Narrative reviews: a less formal review that might include commentary and expert insight. 
    \end{itemize}

In the following sub-sections we provide brief overviews of these three categories of review paper. 

\subsection{Systematic Reviews (SR)}

Systematic reviews emerged at the end of the twentieth century with the objective of establishing a very explicit methodology for conducting reviews \cite{davies2000relevance,greenhalgh2018time}. A systematic review should begin with a focused research question.  Then, the methodology defines how to identify relevant primary research studies with explicit inclusion and exclusion criteria to determine what is considered \cite{davies2000relevance}. The review will synthesise the research findings providing an evidence base for decisions. It may also identify knowledge gaps for future primary research. It would not be unusual in medical research for a systematic review to answer a specific question about how best to treat a specific disease \cite{greenhalgh2018time}. Because systematic reviews follow an explicit methodology, the work should be reproducible; a different research team should produce more or less the same findings. 

The added rigor of systematic reviews means that they enjoy a high status as academic publications, oten representing the top of the publication hierarchy in fields such as medicine \cite{greenhalgh2018time}. With this status comes the potential to attract citations, a factor that has probably contributed to the increase in the proportion of systematic reviews in the research literature in recent years. 

In 2016 Ioannidis published an influential criticism of the state of systematic reviews and meta-analyses in medical research \cite{ioannidis2016mass}. The main criticism was that the relative ease with which these case be produced resulted in a huge increase in publications. There has been a greater that 25-fold increase in systematic reviews and meta-analyses in PubMed between 1986 and 2014, a period in which the volume of original research papers increased by a little over 50\%. Ioannidis argues that there is massive redundancy in this work with some meta-analyses being repeated over 20 times. 
A separate analysis of a sample of papers from PubMed form 2000 to 2019 by Hoffmann \emph{et al.} \cite{hoffmann2021nearly}
suggests that there has been a 20-fold increase in systematic reviews in that time. This compares with a 2.6-fold increase in PubMed publications overall. The authors argue that there is significant overlap in these systematic reviews resulting is considerable research waste.

\subsection{Meta-Analyses (MA)}

In some circumstances it will be possible to take the work in a systematic review a step further and provide an integrated quantitative analysis. This meta-analysis is a statistical process that combines the quantitative results from different primary studies \cite{davies2000relevance}. In order  to combine  results from the primary studies they need to share a common methodology. If the results can be combined it will provide a more accurate estimate of effect size and should increase the statistical power to detect significance. If the primary studies do not have similar methodologies it will not be possible to do a meaningful meta-analysis. 

It will almost always be the case that a meta-analysis follows on from a systematic review that identifies the primary research studies that are to be integrated. However, it is possible that the meta-analysis integrates studies that have been identified through an informal process; in which case there is the risk that the analysis is subject to a selection bias. 

It is in the nature of the research questions being considered in some FoS that meta-analysis are not always possible or appropriate. This is true for Computer Science and Physics for instance. However, meta-analyses are commonplace in Psychology and in Medicine where quantitative studies are amenable to meta-analysis. Research by Patsopoulos \emph{et al.} from 2005 \cite{patsopoulos2005relative} showed that meta-analysis papers have the highest citation impact among publications in the Health Sciences. 

\subsection{Narrative Reviews (NR)}
Narrative reviews are sometimes referred to as `traditional' reviews. Their characteristics  are not as clearly defined as systematic reviews \cite{ferrari2015writing}; in fact they are sometimes defined by what they are not--they are not systematic reviews \cite{davies2000relevance}. While narrative reviews may lack the rigor and reproducibility of systematic reviews, they do have some advantages:
\begin{itemize}
    \item They may be appropriate for bringing together results of studies that have employed diverse methodologies.
    \item Narrative reviews can provide a historical account of the development of theory and research on a topic \cite{ferrari2015writing}
    \item They provide  room for subjective insights and conjectures about future directions in the field of research \cite{gupta2023chatgpt}.

\end{itemize}

The other claimed benefit for narrative reviews is that they typically draw on expert opinion by deliberately recruiting
leading names in the field \cite{greenhalgh2018time}. So while systematic reviews and meta-analyses attract more citations on average, good narrative reviews have the potential to be very influential. The list of 50 most cited reviews of 2013-2017 \cite{bagirova202150} contains many narrative reviews. In fact the paper on Deep Learning in first place would be considered a narrative review \cite{lecun2015deep}.

\section{The Use of AI in Academic Paper Writing}\label{sec:AI_Use}
Since the launch of ChatGPT in 2022, the use of GenAI tools in the writing of research papers has recived a lot of attention from researchers \cite{wu2025survey,bao2025there,liang2024mapping,cunningham2025analysis}. The editorial policy for most journals allows the use of AI for language improvement but a broader authorship role is not permitted. 
As GenAI tools evolve it will become more and more difficult to detect their use in scientific writing \cite{liu2019roberta}. Liang \emph{et al.} \cite{liang2024mapping}
produce estimates of the extent of AI use in papers published up to 2024. They find the most use in Computer Science at 17.5\% and the least in Mathematics at 5\%. The assertion is that these percentages 
reflect the proportion of sentences `substantially modified by LLM' in abstracts published on arXiv. Their analysis showed that these percentages were on the rise since 2023 so it is reasonable to assume that AI use may be even more prevalent now. 

Given that the production of review papers is a research activity that is all document-based, it is to be expected that GenAI will have more impact there than in the wider scientific literature. While software tools to support the systematic review process have been around for some time \cite{kohl2018online,oconnor2024large}, the arrival of GenAI in 2023 presents a step change in the automation that is possible \cite{bolanos2024artificial,qureshi2023chatgpt}.
Qureshi \emph{et al.} \cite{qureshi2023chatgpt} present an analysis of the potential for GenAI to contribute at each step in the review process. In their analysis, the main stages are: formulating the research question, conducting the document search, synthesising and summarising the material and producing the code for meta-analysis if required. They found that  ChatGPT3.5 was helpful in summarising the material and producing code for meta-analysis. It was less useful for problem formulation and conducting a comprehensive literature search where fabrication proved to be a significant problem. 

So, how much of the work is preparing a review paper can be done by GenAI tools? Bolanos \emph{et al.} suggest that before the end of the decade, AI-enabled research assistants will be available that will be able to generate comprehensive literature reviews. Indeed there are reports that this time has already arrived. Qiu \emph{et al.} \cite{qiu2025completing} presents \emph{Insight}Agent, a human-in-the-loop tool for automating the production of systematic reviews. They claim that their tool can produce a high quality review that would previously have taken months of work in a few hours. In \cite{cao2025automation}
the authors claim that they reproduced 12 Cochrane reviews in two days, representing 12 works-years of traditional systematic review work. 

While the quality of these high-speed reviews must be open to question, it is clear that the arrival of GenAI dramatically changes  the literature review landscape. Yun \emph{et al.} propose that that arrival of LLMs for producing literature reviews opens the prospect of on-demand synposes on a topic of interest--literature reviews will be automated and on-demand \cite{yun2023appraising}.

\section{Materials and Methods}
In this section, we describe the collection and preparation of a dataset of almost 17 million scientific article records from Semantic Scholar (\url{semanticscholar.org}) covering the period 2000 through 2024. We describe the key metrics and statistical methods used in this work.

\subsection{Dataset}
The data used in this work was collected from the Semantic Scholar during May 2025. Semantic Scholar was chosen because it is an established dataset for scientometrics research \cite{cunningham2024analysis,he2025academic,10.1145/3742442,liang2021finding} that is well-suited for the present study, and its open-source nature offers significant accessibility and reproducibility benefits \cite{rothenberger2021mapping} compared with proprietary Web of Science and Scopus alternatives, which are only available under license \cite{kinney2023semantic}.
The data collected includes articles published from 2000, up to and including 2025 (partial year), and whose meta-data included \emph{'JournalArticle'} in the \emph{'publicationtypes'} field. This produced an initial dataset of 29,893,335 articles, the 30M dataset.

\subsubsection{The 17M Dataset}
Not all Sematic Scholar records are complete \cite{delgado2024completeness} and to ensure we had sufficient data for the present study, we implemented the following filtering rules:

\begin{enumerate}
    \item Exclude articles where abstracts had less than 5 words (35\% of articles)
    \item Exclude articles with no references (23\% of articles).
    \item Exclude articles published after 2024; 2025 was a partial year at the time of writing.
\end{enumerate}

These rules reduced the 30M dataset to 16,699,646 articles, which we refer to as the \emph{17M dataset}. We use this 17M dataset exclusively in the analysis that follows and it is summarised in Appendix \ref{app:dataset_summary}.

We are interested in analysing articles across different fields of study. Semantic Scholar records include an ordered list of \emph{fields of study} (FoS), containing at least one FoS per article. We used the initial field in this list as the article's \emph{primary field of study} -- identifying 23 different primary FoS -- to separate the articles into 23 mutually exclusive sets, each corresponding to a different primary FoS.

\subsubsection{Categorising Review Papers}
Semantic Scholar's \emph{publicationtypes} field can also contain \emph{Review} and \emph{MetaAnalysis} values, which, in theory, help to identify review papers. In practice, however, these identifiers are not sufficient to produce a reliable classification of review papers. We address this in two ways:

\begin{enumerate}
    \item We look for certain key phrases in the title or abstracts of papers that are likely to be reliable indicators of review papers; see Appendix \ref{app:review_phrases}. 
    \item We identify venues or journals that only publish review papers. These so-called \emph{review venues} are venues that include \emph{`review', `survey', or `trends in'} in their titles, such as \emph{Computing Surveys} or \emph{Trends in Cell Biology}. 
\end{enumerate}

    A paper is considered to be a \emph{review} paper, for the purpose of this study, if \emph{Review} or \emph{MetaAnalysis} is included in its \emph{publicationtypes} field \emph{and}, either its title or abstract includes one of the strict review phrases \emph{or} it appears in a \emph{review venue}. Papers that are not review papers are refereed to as \emph{regular} papers.
    
    Finally, we identify the different types of review papers as follows:
    
\begin{enumerate}
    \item A review paper is a meta-analysis review (MA) if it's \emph{publicationtypes} field contains \emph{MetaAnalysis} or it's title or abstract contains a strict review phrase which itself contains the keyword \emph{`meta'}.
    \item A review paper is a systematic review (SR) if it is not a meta-analysis review and if its title or abstract contains a strict review phrase which itself contains the keyword \emph{`systematic'}.
    \item A review paper is a \emph{narrative review} (NR) if it is neither an MA nor an SR.
\end{enumerate}

In the 17M dataset there are 642,689 review papers (3.8\% or all papers), made up of 111,395 (17.3\%) MAs, 138,260 (21.5\%) SRs, and 393,034 (61.2\%) NRs.

\subsection{Key Metrics}
The main analysis will focus on the citation impact of different types of papers, and how this has changed over time. We will also examine the evidence for recent AI usage across fields of study and whether this varies with paper type (regular vs. review).

\subsubsection{Quantifying Citation Impact}
For citation impact, we account for temporal and field of study effects, by using the \emph{normalised citation index} \cite{Kenna2017scientometrics}, which divides the citation count of a paper $i$ (published in year $y$ and field $f$) by the mean citation count for papers published in the same field and year; see Equation \ref{eq:nci}. This single-field NCI metric was chosen because of our focus on a single, primary field of study for each publication.

\begin{equation}
\mathrm{NCI}_{i} = \frac{C_{i}}{\overline{C}_{f,y}}
\label{eq:nci}
\end{equation}



\subsubsection{Quantifying AI Use}
To quantify the use of modern AI tools in scientific writing we focus on the subset of papers published in 2024, the first full year after the launch of ChatGPT and the modern AI era. To do this we use a term-based AI scoring strategy using a set of terms most commonly associated with AI texts provided by Liang \emph{et al.} \cite{liang2024monitoring, liang2024mapping}. We use odds ratios to identify the top-300 most over-represented terms (\emph{AI amplified terms}) from the AI-generated texts; the top 10 AI and human terms are listed in Table \ref{tab:top_terms}. While the human terms are not used in our scoring, they are included here to provide a sense of the difference in tone between AI and human text.  For each paper we count the total occurrences of these AI amplified terms in the abstract and calculate a \emph{raw AI score} as this count per 100 abstract words.


\begin{table}[h]
\caption{The top 10 AI and human terms in the training data as scored using Odds Ratio.}\label{tab:top_terms} \label{tab:top_terms}
\includegraphics[width=\textwidth]{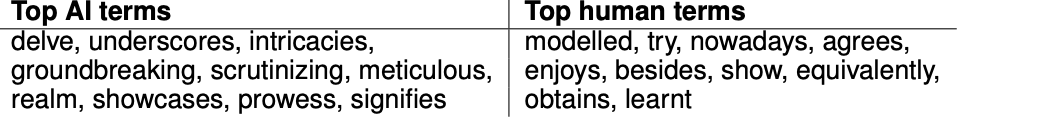}
\end{table}


To produce a field-normalised version of this raw AI score, we calculate a \emph{baseline AI score} for each primary field of study from the mean raw AI score for 2020 articles in that field; 2020 was chosen as a suitable recent year prior to the appearance of mainstream AI tools. This baseline represents the \emph{ambient} use of AI amplified terms before the widespread adoption of AI tools. Then, we calculate an \emph{excess AI score} for each 2024 article, by subtracting the corresponding baseline AI score from the article's raw AI score. We use this excess AI score to compare the strength of the AI signal among different paper types in 2024.




\subsection{Statistical Methods}
The main statistical analysis in this paper focused on a comparison of citation impact and the strength of the AI signal across different fields, years, and paper types. In all cases, medians were used as robust indicators of central tendency, due to the skewed nature of citation and AI signal data; 95\% confidence intervals were estimated using the Hettmansperger-Sheather method. The analysis was conducted in Python (v3.12) using the \texttt{SciPy}, \texttt{statsmodels}, and \texttt{scikit-posthocs} libraries.

\subsubsection{Comparing Citation Impact}
Differences in NCI by paper type (2000-2024) were evaluated using non-parametric methods; given the large sample sizes and the skewed distribution of citation data, normality was not assumed. Group-level differences were tested using the Kruskal--Wallis H test, as a rank-based alternative to one-way ANOVA. Where significant heterogeneity was detected, post-hoc pairwise comparisons were performed using Dunn’s tests with Bonferroni correction for multiple testing. Effect sizes were expressed as \(\varepsilon^{2}\) (epsilon squared).

We used a similar approach to evaluate differences in NCI during two 10-year periods, 2000-2009 and 2015-2024, \emph{before} and \emph{after} the introduction of several tools to support the production of review papers; see \cite{aria2017bibliometrix, Cleo2019Usability, marshall2015systematic, nuijten2016statcheck, oouzzani2016rayyan,eppiReviewer2018,vandeschoot2020asreview,distillersrPrioritization2020,wallace2009metaanalyst}. Then, we compared changes in NCI between these periods using the Mann--Whitney U test; the strength and direction of between-period differences were quantified using Cliff’s delta and multiple testing adjustments were applied using the Bonferroni correction, with statistical significance set at \(\alpha = 0.01\). 


\subsubsection{Comparing AI Scores}
To evaluate the strength on the AI signal across fields we focused on the 2024 papers in the top quartile per FoS by excess AI score; these are the recent papers in each field with the strongest AI signal. Differences in excess AI score distributions across FoS were assessed using a Kruskal–Wallis H-test with Bonferroni-adjusted Dunn’s post-hoc pairwise comparisons (\(\alpha = 0.01\)). Pairwise effect sizes were computed using Cliff’s delta. Median AI-scores and mean absolute \(|\delta|\) values were used to quantify and rank fields by the AI signal strength.

To test whether review papers exhibited larger excess AI scores than regular papers, we compared the excess AI score distributions
using non-parametric Mann–Whitney U tests. Analyses were performed separately within each FoS to control for disciplinary heterogeneity, with Cliff’s delta (\(\delta\)) used to estimate effect sizes. The p-values were Bonferroni-adjusted for multiple comparisons. In addition to field-level tests, we conducted a pooled Mann–Whitney U test (ignoring field structure) and a stratified analysis combining per-field
z-scores using Stouffer’s weighted-Z method, which yields an overall field-adjusted p-value indicating whether within-field effects are directionally consistent across disciplines.

As with citation impact, differences in the excess AI score by paper type were also evaluated, using the Kruskal--Wallis H test followed by post-hoc pairwise comparisons using Dunn’s tests with Bonferroni correction; effect sizes were expressed as \(\varepsilon^{2}\).

\section{Results}

\subsection{The Proportion and Growth of Reviews}

\begin{figure}[!ht]
    \centering \includegraphics[width=\textwidth]{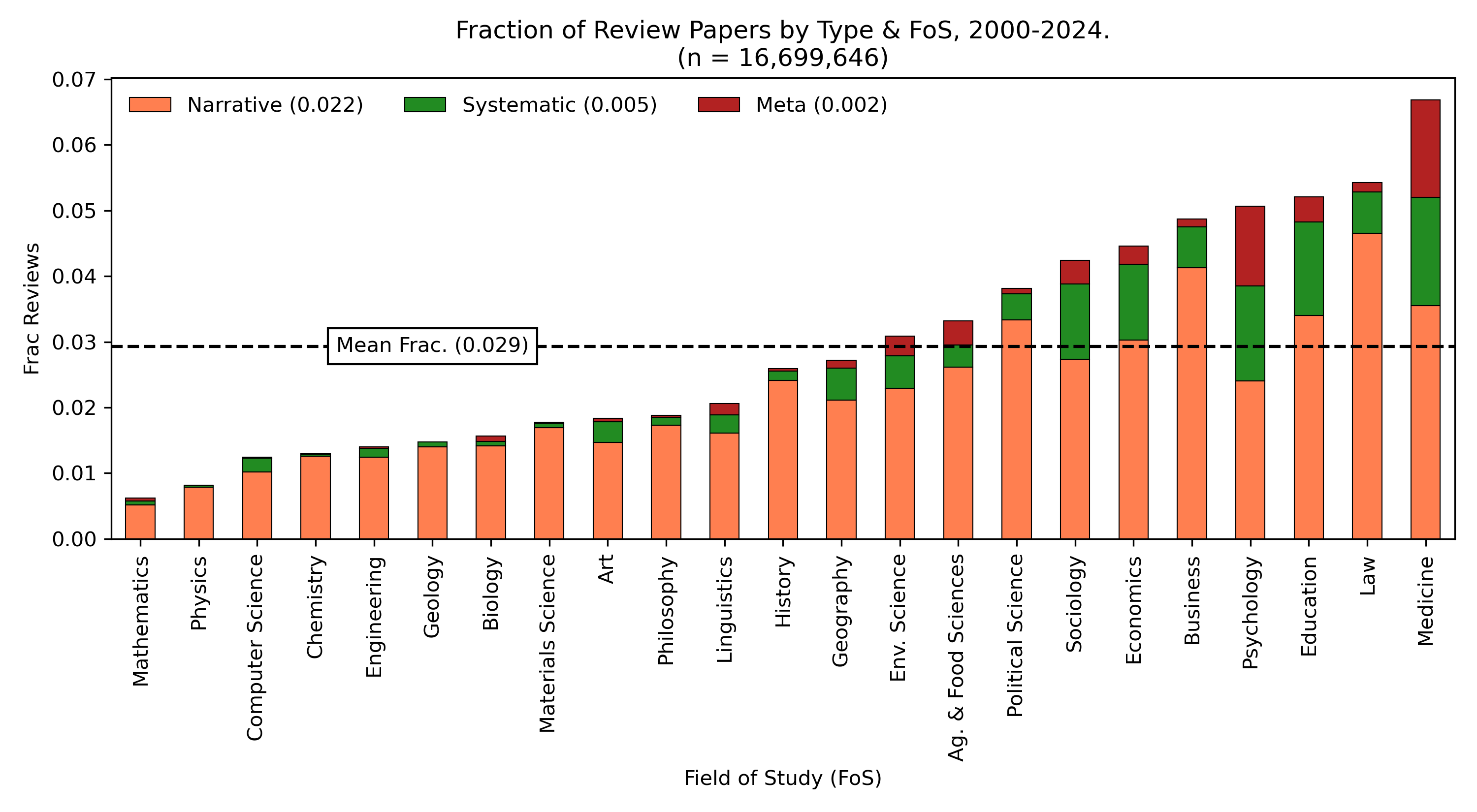}
    \caption{Proportion of papers by review type and field of study and FoS. The fields of study are sorted in ascending order of the overall fraction of review papers. The dashed horizontal line indicates the overall fraction of reviews, averaged across all 23 fields of study, and the numbers in brackets in the legend indicate the macro-average of the fraction of different types of reviews for different fields of study.}
    \label{fig:RP_frac_FiS}
\end{figure}

Figure \ref{fig:RP_frac_FiS} shows a stacked bar-chart with a count of different types of review papers expressed as a fraction of the total number of papers in each field of study. The overall height of the bars corresponds to the overall fraction of review papers for each field of study. 

\begin{figure}[!ht]
    \centering \includegraphics[width=\textwidth]{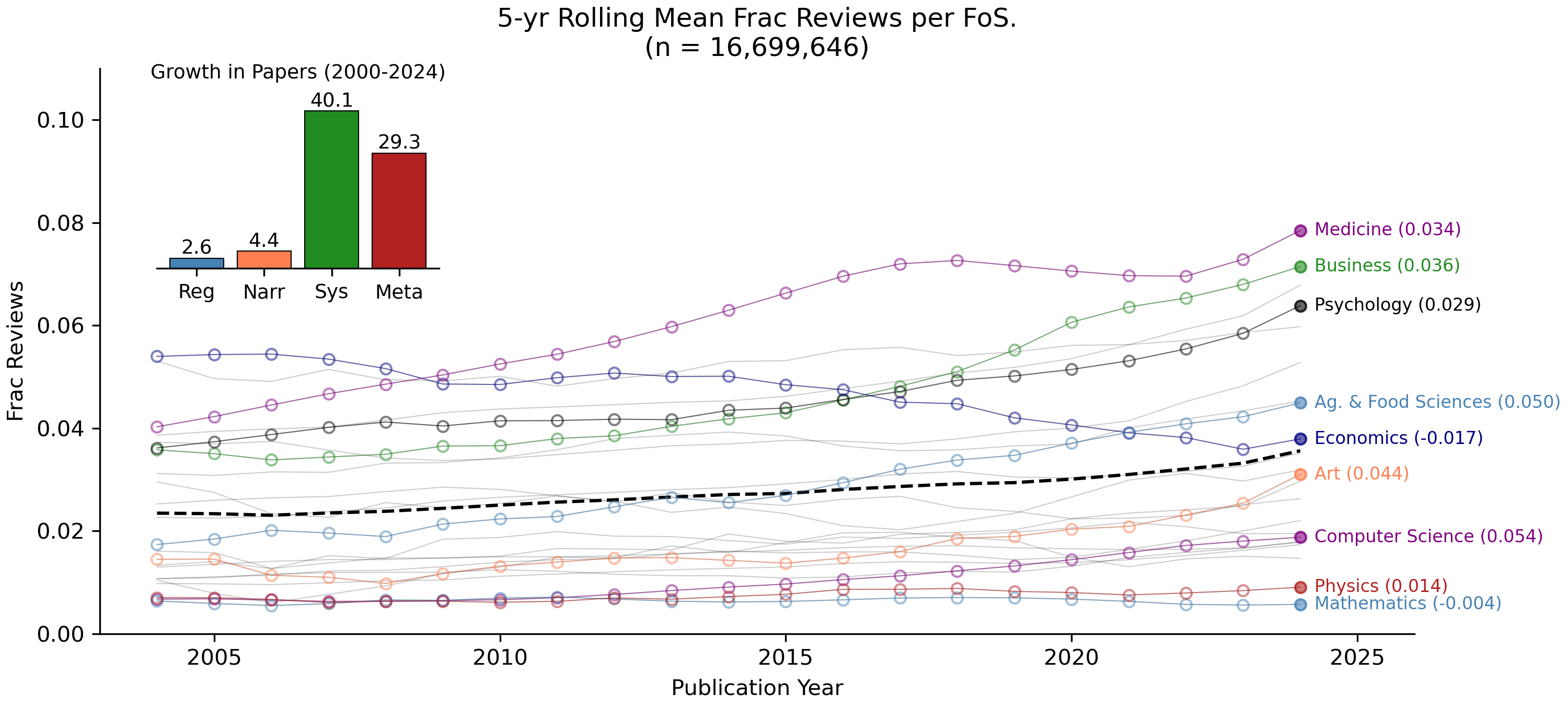}
    \caption{The overall proportion of review papers by FoS over time. The black dashed line indicates the macro-average of the fraction of reviews across all FoS. The numbers in brackets after each highlighted field name corresponds to the average year on year growth rate for that field. The inset bar-chart shows the total growth in papers between 2000 and 2024.}
    \label{fig:RP_frac_Time}
\end{figure}

Figure \ref{fig:RP_frac_Time} shows this overall fraction of reviews, for each FoS, during the period 2000-2024, based on a five-year rolling average. A subset of FoS are highlighted and labeled. The inset bar-chart  shows the total growth in papers between 2000 and 2024. During this time-frame there was a 2.6-fold and a 4.4-fold increase in regular papers and NRs, respectively, compared with a 40.1 and 29.3-fold increases in SRs and MAs.




\subsection{The Citation Impact of Regular \& Review Papers}
\begin{figure}[h]
    \centering \includegraphics[width=\textwidth]{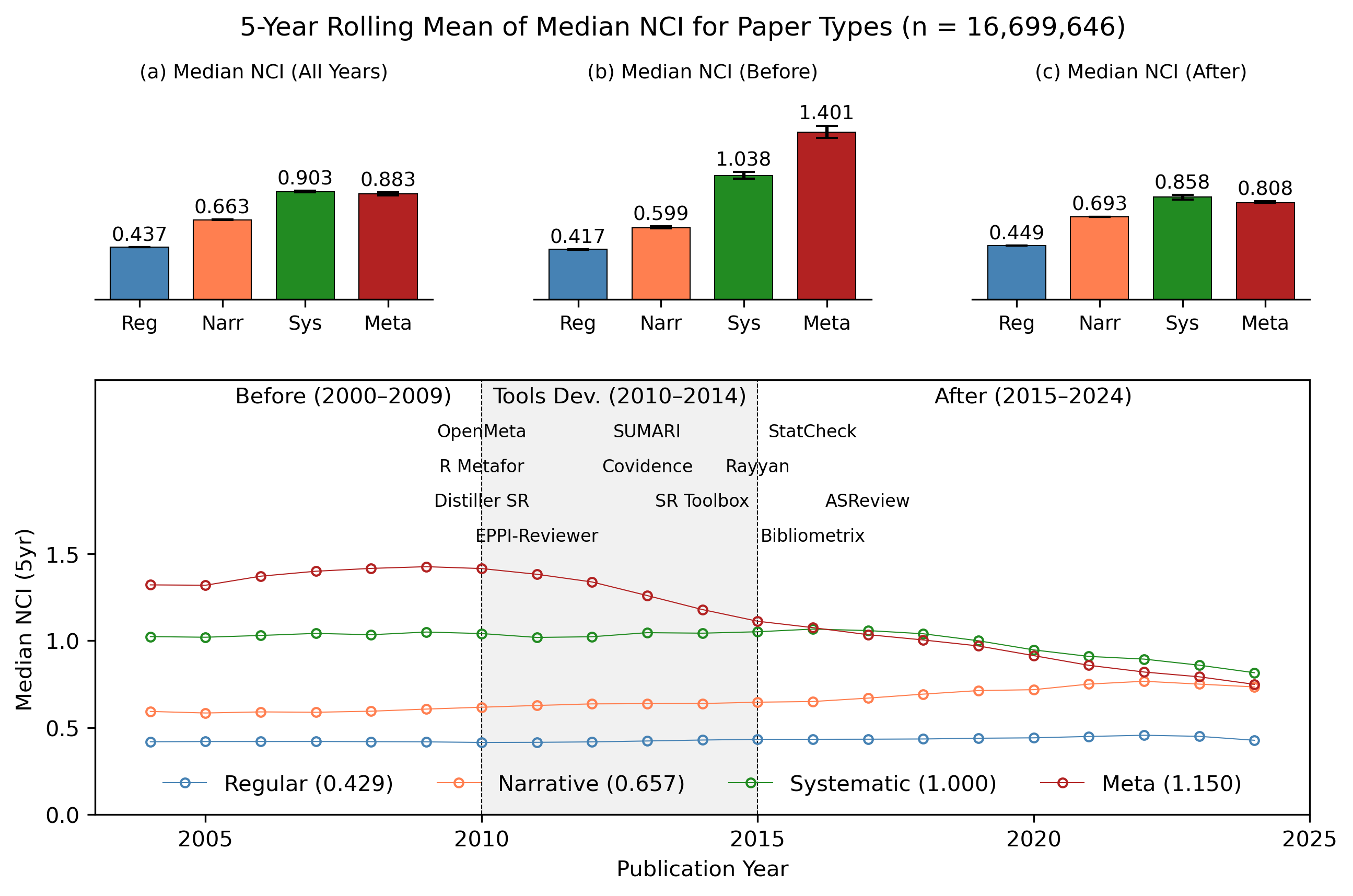}
    \caption{The main line-graph shows the Five‐year rolling median normalised citation index (NCI) by paper type (2000–2024). The bar-charts compare median NCI (with error-bars representing the 95\% confidence intervals) across types for all years (a) and also for two periods (b) before and (c) after the introduction of several major software tools supporting the production of systematic and meta‐analytic reviews (annotated along the approximate timeline, e.g., \textit{Distiller} \cite{distillersrPrioritization2020}, \textit{Covidence} \cite{Cleo2019Usability}, \textit{Rayyan} \cite{oouzzani2016rayyan}, \textit{ASReview} \cite{vandeschoot2020asreview}).}
\label{fig:NCI_Fos_Paper_Type}
\end{figure}

The main line-graph in Figure \ref{fig:NCI_Fos_Paper_Type} shows the five-year rolling average of median NCI for each type of paper from 2000 through 2024. The separate bar-charts show the median NCI values with 95\% CIs for each paper type across three periods: (a) all years, 2000-2024; (b) the ten-year period (2000-2009) before the arrival of several mainstream tools to support the production of review papers; and (c) the ten-year period (2015-2024) after the arrival of these tools.

\subsubsection{Median NCI for All Years, 2000-2024}
From 2000 through 2024, Figure \ref{fig:NCI_Fos_Paper_Type}(a), NCI values differed across paper types, with median NCI values of \(0.437~[0.437, 0.438]\) for regular papers, \(0.663~[0.663, 0.667]\) for NRs, , \(0.903~[0.898, 0.911]\) for SRs, and \(0.883~[0.872, 0.895]\) for MAs. The Kruskal--Wallis test indicated a highly significant overall difference among groups (\( H = 99{,}688.18 \), \( p < 0.001 \)), with a small effect size (\( \varepsilon^{2} = 0.006 \)). Bonferroni-adjusted Dunn’s post-hoc tests revealed that review papers (NRs, SRs, and MAs) all had significantly higher NCI values than regular papers (\( p < 0.001 \)), and that both SRs and MAs also exceeded NRs (\( p < 0.001 \)). No significant difference was observed between SRs and MAs. Further statistical results are provided in Appendix~\ref{app:nci_2000_2024}.


\subsubsection{Comparing Before (2000-2009) \& After (2015-2024) Periods}
For the before and after periods, Figures \ref{fig:NCI_Fos_Paper_Type}(b) and (c), Kruskal--Wallis tests also confirmed significant heterogeneity among groups, \( H = 15{,}496.92\) and \(64{,}301.83\), respectively with \( p < 0.001 \). And for both periods the Bonferroni-adjusted Dunn’s post-hoc tests also confirmed that NRs, SRs, and MAs had significantly higher NCI values than regular papers (\( p < 0.001 \)). Further statistical results are provided in Appendix~\ref{app:nci_2000_2009} and \ref{app:nci_2015_2024}.  

However, Figures \ref{fig:NCI_Fos_Paper_Type}(b) and (c), also show an important shift in NCI values between these periods. A Bonferroni-adjusted, Mann--Whitney U test revealed statistically significant changes across all paper types (\(p < 0.001\)). MAs showed the largest decline, with median NCI decreasing from 1.401 to 0.808 (\(U = 4.7 \times 10^{8}\)). SRs also exhibited a decline (\(1.038 \rightarrow 0.858\); \(U = 6.6 \times 10^{8}\)). In contrast, narrative reviews and regular papers displayed more modest but still significant improvements in median NCI, from 0.599 to 0.693 (\(U = 9.5 \times 10^{9}\)) and from 0.417 to 0.449 (\(U = 1.7 \times 10^{13}\)), respectively. Further statistical results are provided in Appendix~\ref{app:nci_beween_periods}.

\subsection{Comparing the AI Signal by Field of Study and Paper Type}

\subsubsection{The AI Signal by Field of Study}

\begin{figure}[!ht]
    \centering \includegraphics[width=\textwidth]{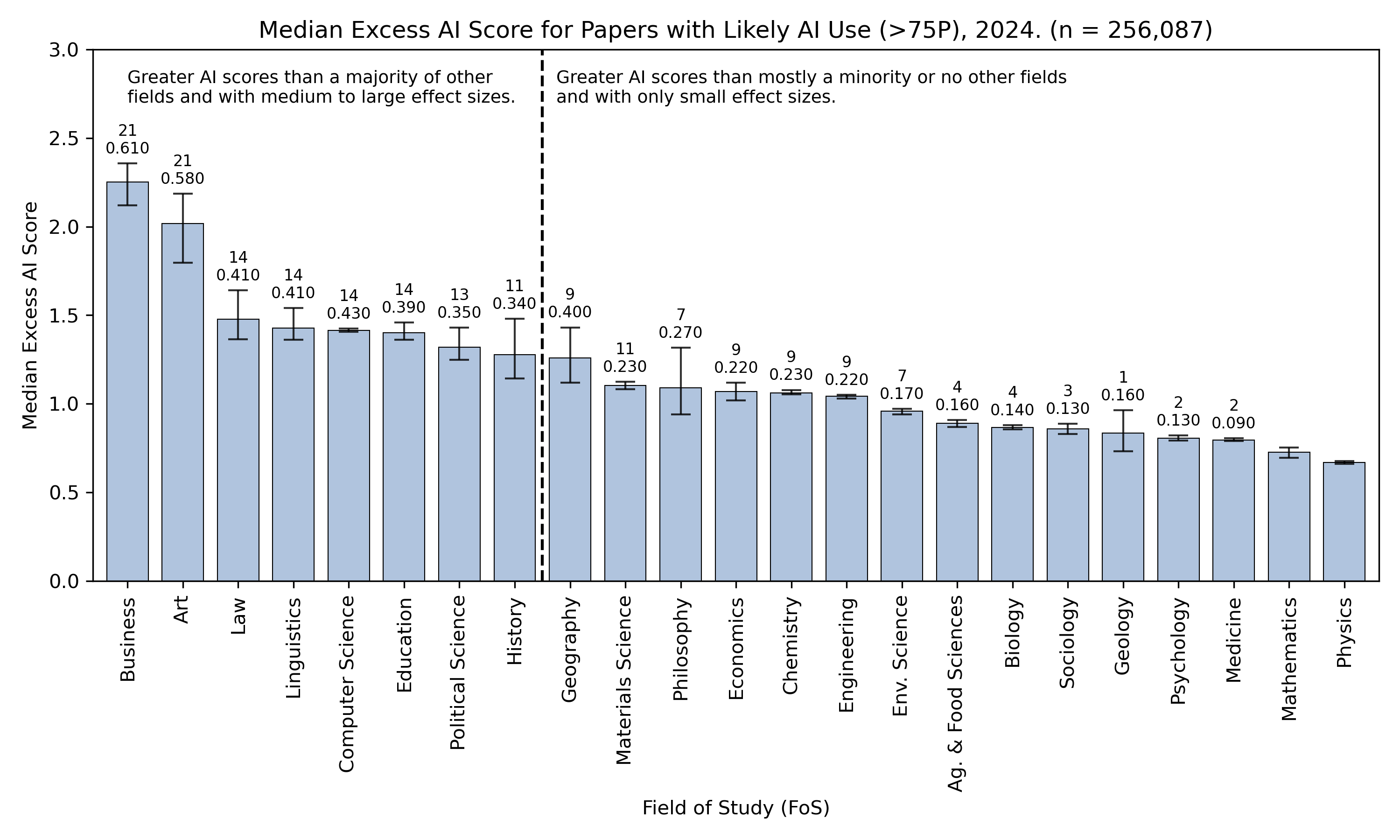}
    \caption{Median excess AI-score (with error-bars representing the 95\% confidence intervals) by field of study for papers with likely AI usage in 2024. Results are based on Kruskal–Wallis (\(H =33{,}256.2\), \(p < 0.001\), \(\epsilon^2 = 0.130\)) and Bonferroni-corrected Dunn’s post-hoc tests with pairwise effect sizes estimated via Cliff’s delta. Numerical labels above each bar indicate the number of other fields with significantly lower AI-scores and the mean absolute Cliff’s delta for those contrasts. The dashed line separates fields with AI-scores greater than a majority of other disciplines with medium to large average effect sizes (left) from those exceeding only a few or none (right).}
\label{fig:excess_ai_by_fos}
\end{figure}

Figure \ref{fig:excess_ai_by_fos} shows the median excess AI score for 2024 papers in the top-quartile of each of the 23 fields. The Kruskal–Wallis test revealed significant variation in median AI-scores
across these fields (\(H = 33{,}256.2\), \(p < 0.001\)), indicating systematic field-level differences of medium (\(\epsilon^2 = 0.130\)) effect size. Post-hoc Dunn’s tests (Bonferroni-corrected, \(\alpha = 0.01\)) identified
199 significant pairwise contrasts. Across all comparisons, the average absolute Cliff’s delta was \(|\delta|_{\text{mean}} = 0.307\)
(\(|\delta|_{\text{median}} = 0.278\)), reflecting consistent small-to-moderate between-field effects; see also Appendix~\ref{app:ai_score_by_fos}

\subsubsection{The AI Signal for Regular \& Review Papers}
\begin{figure}[!ht]
    \centering \includegraphics[width=\textwidth]{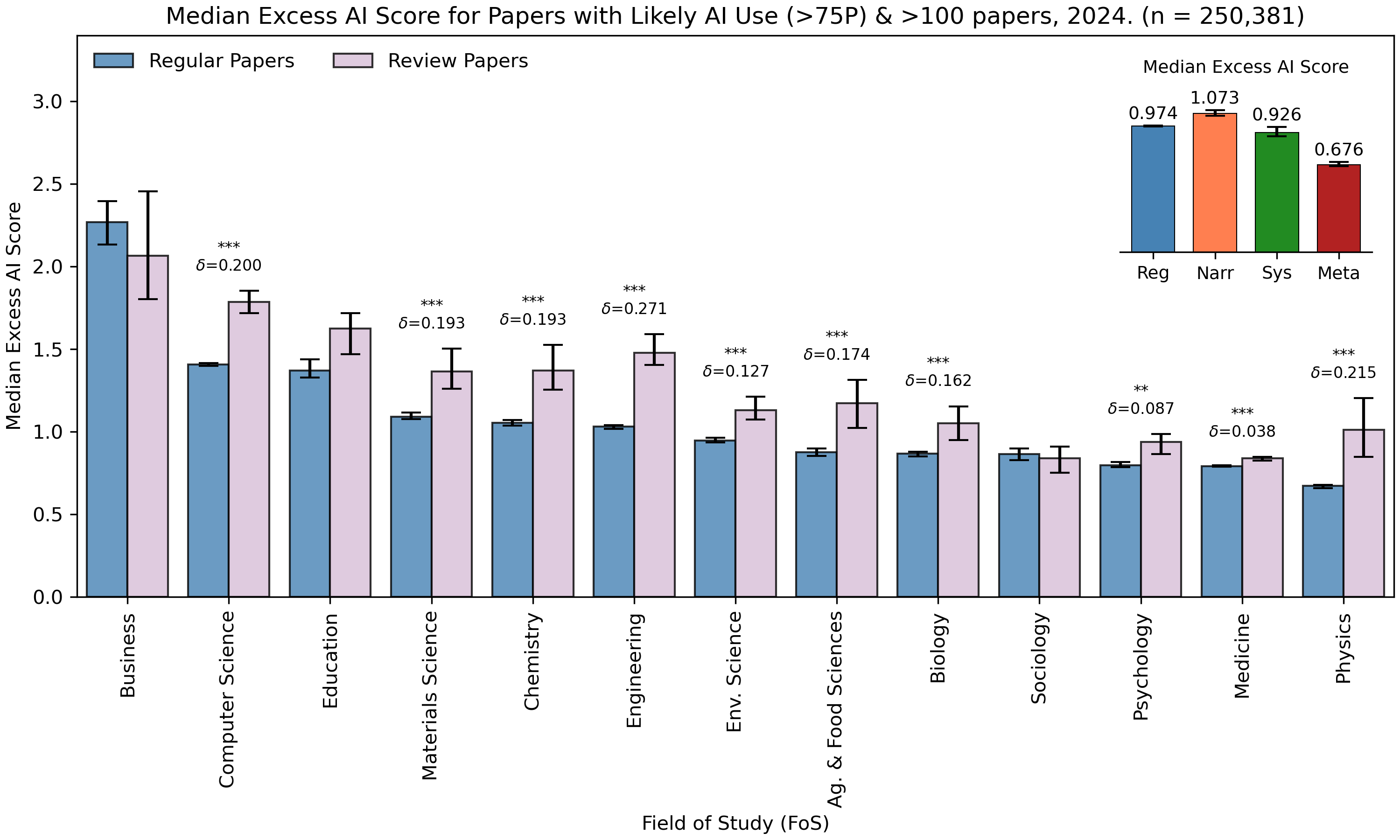}
    \caption{The main bar-chart shows the median excess AI-scores by FoS (with error-bars representing the 95\% confidence intervals) for review and regular papers with likely AI use (>$75^{\text{th}}$ percentile) with at least 100 papers in 2024, \(n = 250{,}381\)). Bars are labeled with significance levels (*, **, ***) denoting Bonferroni-corrected Mann–Whitney tests comparing review vs. regular papers within each field; Cliff's \(\delta\) effect sizes are displayed numerically above bars for fields with significant ($p<0.01$) differences. The smaller inset bar-chart shows the median excess AI score (with error-bars representing the 95\% confidence intervals) for each type of paper across all 13 FoS.}
\label{fig:median_ai_score_by_fos_likely_ai_reg_vs_rev}
\end{figure}

Figure \ref{fig:median_ai_score_by_fos_likely_ai_reg_vs_rev} shows the median excess AI score for regular and review papers in each FoS during 2024, again for the top quartile of papers per field by excess AI score, but excluding (10) fields with fewer than 100 review papers in 2024. For the remaining 13 fields, review papers exhibited higher AI scores than
regular papers in aggregate. The pooled Mann--Whitney U test indicated a small
but statistically significant overall difference (\(U = 2.1 \times 10^{9}\),
$p=0.044$), with median AI scores \(0.984~[0.967, 0.999]\) and
\(0.974~[0.972, 0.979]\) for reviews and regular papers, respectively; see Appendix~\ref{app:ai_score_by_fos_reg_vs_rev_mann_whitney}.

When controlling for field of study, the stratified (FoS-adjusted) Mann--Whitney test yielded a highly significant combined result (\(Z = 21.31\), \(p < 10^{-100}\)), confirming a consistent within-field pattern across the fields (mean \(\delta = 0.130\); median \(\delta = 0.162\)). Field-level analyses showed significant differences in 10 of 13 fields, with review papers associated with higher median AI scores in all 10 cases;  see also Appendix~\ref{app:ai_score_by_fos_reg_vs_rev_mann_whitney}.

The insert bar-chart in Figure \ref{fig:median_ai_score_by_fos_likely_ai_reg_vs_rev} compares the median AI scores for each type of paper in this set of papers. The Kruskal--Wallis test revealed a statistically significant overall difference
among groups (\(H = 661.64\), \(p < 10^{-140}\), \(\varepsilon^{2} = 0.003\), small effect) and Post-hoc Dunn tests with Bonferroni correction confirmed all pairwise comparisons were significant (\(p<0.01\)), with NRs exhibiting the highest
median AI score (\(1.073~[1.053, 1.098]\)), followed by regular papers (\(0.974~[0.972, 0.979]\)), SRs (\(0.926~[0.896, 0.967]\)), and MAs (\(0.676~[0.663, 0.698]\)); see also Appendix~\ref{app:excess ai score_2024_2024}

\section{Discussion}

\subsection{On the Growth of Review Papers}

Figure \ref{fig:RP_frac_FiS} shows that review papers accounted for approximately 3\% of all papers written across the 23 fields of study in our dataset between 2000 and 2024, but with considerable variation by field. Reviews are most common (almost 7\%) in Medicine but they are rare (<1\%) in Mathematics and Physics. NRs are the most common (2.2\% of all papers) type of reviews across all fields of study, while MAs are the least common (0.2\%), and mostly evident only in clinical or human-study fields, such as Medicine, Psychology, Sociology, and Agricultural and Food Sciences.

The fraction of review papers is growing in most fields; see Figure \ref{fig:RP_frac_Time}. They grew fastest in Computer Science, Agricultural and Food Sciences, and Art; approx. 4-5\% per year. Mathematics and Economics were the only fields where the fraction of reviews fell over the period, although for the latter, it is worth noting that review papers in Business have been growing at $>$3\% per year, and perhaps some Economics papers are being classified as Business papers.

The bar-chart in Figure \ref{fig:RP_frac_Time} summarises the total growth in different types of papers from 2000 to 2024. SRs and MAs increased by a factor of 40.1 and 29.3, respectively, compared to more modest growth of 2.6 and 4.4 for regular papers and NRs. This is consistent with the findings of Ioannidis~\cite{ioannidis2016mass}, albeit over the period from 1991 to 2014: PubMed SRs and MAs grew by 27x and 26x, respectively, compared to just 1.5x for all PubMed papers. Ioannidis warned that this constituted an alarming over-production of SRs and MAs that would be harmful to research and diminish the value of such papers. Similar concerns, and growth rates, were noted by Hoffmann \emph{et al.}~\cite{hoffmann2021nearly} based on an analysis of more than 1,100 systematic reviews from more than 570,000 PubMed papers, from 2000 through 2019, and which saw a $>$20x increase in the number of SRs. 

\subsection{On the Decline of Review Paper Impact}

Generally speaking reviews are assumed to offer a citation benefit \cite{evans2003hierarchy} and the results in Figure \ref{fig:NCI_Fos_Paper_Type} support this. Overall (2000-2024) review-based papers (NRs, SRs, and MAs) demonstrated markedly higher field-normalised citation impact than regular papers; the Bonferroni-adjusted Dunn’s post-hoc tests showed that NRs, SRs, and MAs all had significantly higher NCI values than regular papers (\( p < 0.001 \)).

Is there any evidence that the over-production of reviews has change this? Figure \ref{fig:NCI_Fos_Paper_Type} shows that the citation benefit for review papers has been declining. SRs and MAs experienced a statistically significant decline in NCI ($p<0.01$) between the periods 2000-2009 and 2015-2014, against a modest but statistically significant increase ($p<0.01$) in the NCI of regular papers and NRs; currently, SRs enjoy the largest citation benefits, rather than MAs as was the case in the past.

If the growth in SRs and MAs is associated with more research-waste, overlap, and duplication then it is not surprising to see this NCI decline. The  availability of tools to support and even partially automate the production of SRs and MAs may also be a factor: reducing the cost to produce an SR or MA may lead to a proliferation of these papers, diluting their impact. Indeed, we can see from Figure \ref{fig:NCI_Fos_Paper_Type} how the decline SR and MA impact began when such tools emerged around 2010 and has continued since.

Interestingly, and somewhat as an aside, although NRs do not enjoy the same citation benefits as SRs and MAs, there is some evidence that, at the extremes of impact, NRs tend to outperform other papers; see \cite{bagirova202150}, where many of the top-50 most cited papers (2013-2017) are NRs. Among the top 1\% of papers by NCI in our dataset, more than 8.3\% are NRs compared to 2.3\% NRs in the full dataset; ~2\% are SRs and MAs, compared with $<$1\% in the full dataset. In other words, there are $>$3.5x as many NRs among the top 1\% papers, by NCI, than we might expect; there are only 2-3x as many SRs and MAs in this top 1\% set compared to the full dataset. Furthermore, these NRs have a median NCI of 13.1 compared to 12.4 and 12.3 for the SRs and MAs, respectively. Moreover, that 13.1 median NCI for the top 1\% NRs is 20x greater than the overall median NCI for NRs; the median NCI scores for the top 1\% SRs and MAs is only 14x greater than their overall median NCIs.

These findings suggest that that review papers do enjoy a citation benefit but over-production is negatively impacting their value. The availability of new tools that target SRs and MAs may be an important factor, because SRs and MAs have experienced the greatest decline in NCI during the period before and after the introduction of these tools.

\subsection{The Evidence for Increased AI Usage in Review Papers}

The results in Figure \ref{fig:excess_ai_by_fos} build on recent work by \cite{bao2025there, liang2024mapping} by demonstrating the presence of an AI signal that varies in strength across fields. Computer Science, one of the larger fields, also had one of the highest AI scores (0.014) exceeding that of 14 other fields and with moderate effect sizes (mean $|\delta|$=0.430). In contrast, other large fields such as Medicine, Biology, and Physics presented with much weaker AI signals. Whether the impact of new AI tools will be almost immediate \cite{qiu2025completing,cao2025automation} or take longer to be felt \cite{bolanos2024artificial}, the results in Figure \ref{fig:median_ai_score_by_fos_likely_ai_reg_vs_rev} indicate that review papers are already attracting greater AI use than regular papers. Most of the 13 fields analysed showed a stronger AI signal in review papers than regular papers, and the stratified test confirmed this pattern was not field-specific but general, with the strongest effects found for Engineering, Physics, and Computer Science ($|\delta|$=0.20-0.27). 


The overall analysis of AI scores by paper type -- the inset bar-chart in Figure \ref{fig:median_ai_score_by_fos_likely_ai_reg_vs_rev} -- suggests that the additional AI score associated with reviews may be due to NRs, rather than SRs or MAs; NRs had the highest median excess AI score (\(1.073~
[1.053, 1.098]\)), whereas SRs (\(0.926~[0.896, 0.967]\)) and MAs (\(0.676~[0.663, 0.698]\)) were both lower then regular papers (\(0.974~[0.972, 0.979]\)) and NRs. While the first generation of tools primarily targeted SRs and MAs, modern LLM-based AI tools appear better aligned with the production of NRs, which is not surprising given their more narrative focus. This does not, however, mean that AI tools will not impact the production of SRs and MAs in time. 

Bolanos \emph{et al.} suggested that before the end of the decade, AI-enabled research assistants will be available to generate comprehensive literature reviews \cite{bolanos2024artificial}. Indeed there are reports that this time has already arrived. Qiu \emph{et al.} \cite{qiu2025completing} claim their human-in-the-loop AI tool can produce a high quality SR in only a few hours that would previously have taken months of work. Cao \emph{et al.} claim they reproduced 12 Cochrane reviews in two days, using LLMs, representing 12 work-years of traditional SR effort~\cite{cao2025automation}. When such tools become more commonplace they may render the previous (first) generation of tool, from 2010-2015, obsolete and usher in a new era of even more rapid review production.

As modern AI continues to disrupt the practice of research, these findings show that an AI signal can already be detected across many different fields, with review papers serving as a leading indicator of AI use. If past is prologue, this may lead to an even greater level of review production, which could further diminish the value and impact of this important type of research paper.

\section{Conclusions}

This study traced the evolution, status, and impact of review papers between 2000 and 2024, using a large-scale scientometric dataset of almost 17 million research articles spanning 23 primary fields of study. Across this period, we observed a dramatic growth in the number and proportion of review papers, a trend evident in nearly every discipline.  Although review papers have long enjoyed a citation advantage our results show eveidence that this dividend has been steadily declining. The likely cause is saturation: as reviews proliferate, the citations they once commanded are diluted.

We found that this decline is particularly pronounced for systematic reviews (SRs) and meta-analyses (MAs), which historically delivered the highest citation impact. The timing of this downturn aligns closely with the emergence of software tools designed to automate and accelerate the production of SRs and MAs specifically. In contrast, narrative reviews (NRs) have retained and even improved their influence, perhaps because they depend more on interpretive synthesis than on procedural automation.


We also considered the potential for a new generation of modern AI tools to further disrupt this state of affairs. We found recent evidence of AI involvement in paper writing in all 23 fields of study. In fields with a significant proportion of reviews, we found review papers usually exhibited a stronger AI signal than regular papers, however, this time NRs were impacted more than SRs or MAs, at least for now.




These findings suggest that we are approaching a pivotal inflection point in the evolution of scientific publishing. If the last generation of automation tools transformed the review landscape -- fueling rapid growth but eroding impact and value -- then the emergence of generative AI foreshadows a far greater disruption to come. As AI systems begin to draft, synthesise, and even interpret research autonomously, the barriers to producing reviews will collapse entirely. The ability to generate bespoke literature summaries on demand may flood the scientific record with low-effort, low-quality syntheses, further accelerating the decline in both the status and impact of review papers. Unless the research community adapts -- by re-defining what constitutes a meaningful synthesis, and by developing stronger norms for transparency and authorship -- we risk entering an era where the ease of production overwhelms the integrity of publication.

\section*{Acknowledgments}
Supported by Research Ireland Ireland through the Insight Research Ireland Centre for Data Analytics (12/RC/2289\_P2). 

\section*{Author Contributions}
BS gathered and cleaned the data. BS and PC performed the analysis. BS and PC drafted and edited the paper.

\bibliographystyle{plainnat}
\bibliography{rev_papers}

\clearpage

\appendix

\section{Appendix}
\label{appendix}

\subsection{17M Dataset Summary}
\label{app:dataset_summary}

A summary of the 17M dataset. Each row corresponds to a specific primary FoS, in descending order of the number of articles associated in that field, and showing the article count, and the mean and standard deviations of the number of authors (\emph{Authors}), page counts (\emph{Pages}), and citation counts (\emph{Cites}). Medicine and Computer Science account for more than half of the total dataset, whereas Geography and Geology, the smallest fields, have just over 25,000 articles between them. In this work, we do not make claims about the relative size of different fields of research based on these numbers, because relative sizes of fields may be be greatly influenced by Semantic Scholar's data collection process.

\begin{table}[!ht]
    \centering
        \caption{Summary statistics for the 17M dataset by field of study including the number of papers and the mean and standard deviation of author, page, and citation counts; table rows in descending order of the total number of papers per field.}
\begin{tabular}{lllll}
\toprule
 & Papers & Authors & Pages & Cites \\
\midrule
Medicine & 6,568,726 & 6.3±7.9 & 7.9±4.7 & 34.6±134.4 \\
Computer Science & 2,584,230 & 3.6±2.5 & 8.6±6.1 & 28.7±311.9 \\
Biology & 1,837,782 & 5.5±4.3 & 9.7±5.5 & 52.3±183.2 \\
Engineering & 1,255,537 & 4.3±2.8 & 8.5±5.5 & 23.1±82.1 \\
Chemistry & 922,337 & 5.2±3.0 & 7.8±5.9 & 37.9±139.3 \\
Env. Science & 881,021 & 5.6±5.1 & 9.2±6.4 & 36.5±112.0 \\
Psychology & 464,960 & 4.2±3.5 & 11.9±6.9 & 51.3±212.4 \\
Materials Science & 398,465 & 5.7±3.1 & 8.3±5.9 & 39.3±151.5 \\
Physics & 334,309 & 8.7±75.0 & 4.8±6.8 & 30.3±158.8 \\
Ag. \& Food Sciences & 265,681 & 5.6±3.6 & 9.1±5.1 & 32.2±69.0 \\
Education & 233,423 & 3.4±3.2 & 11.6±7.6 & 28.5±99.6 \\
Sociology & 209,553 & 3.4±2.9 & 15.3±7.7 & 37.3±148.5 \\
Mathematics & 191,698 & 2.5±8.3 & 13.7±10.4 & 22.1±187.9 \\
Business & 158,805 & 2.8±1.7 & 12.4±8.4 & 35.7±146.6 \\
Economics & 105,981 & 3.1±2.3 & 14.1±8.9 & 33.6±108.7 \\
Political Science & 70,647 & 2.2±2.2 & 17.1±8.2 & 30.5±87.8 \\
Law & 47,986 & 2.6±2.2 & 12.8±9.5 & 16.7±47.0 \\
History & 45,132 & 2.1±3.5 & 16.2±9.6 & 15.7±77.3 \\
Linguistics & 44,883 & 3.0±3.0 & 13.3±9.3 & 28.5±84.5 \\
Philosophy & 26,782 & 1.5±1.5 & 13.8±8.8 & 22.6±128.0 \\
Art & 26,268 & 2.7±2.4 & 10.4±7.6 & 14.1±89.4 \\
Geography & 14,508 & 3.5±2.9 & 12.8±8.4 & 34.3±81.5 \\
Geology & 10,932 & 5.0±6.0 & 9.4±8.7 & 37.5±111.1 \\
\bottomrule
\end{tabular}

    \label{tab:summary-stats}
\end{table}

\clearpage

\subsection{Review Phrases}
\label{app:review_phrases}

\begin{table}[h]
\caption{Phrases used to identify candidate review papers based on exact matches with titles and abstracts.}\label{tab:review_phrases}
\begin{tabular}{l}
\\
'literature review', 'literature survey', 'literature study', 'literature analysis',\\
'comparative review', 'comparative survey', 'comparative study', 'comparative analysis',\\
'bibliographic review', 'bibliographic survey', 'bibliographic study', 'bibliographic analysis',\\
'bibliometric review', 'bibliometric survey', 'bibliometric study', 'bibliometric analysis',\\
'scientometric review', 'scientometric survey', 'scientometric study', 'scientometric analysis',\\
'systematic review', 'systematic survey', 'systematic study', 'systematic analysis',\\
'systematic literature review', 'systematic literature survey',\\ 'systematic literature study', 'systematic literature analysis',\\
'meta analysis', 'meta review', 'meta study', 'meta-analysis', 'meta-review', 'meta-study',\\
'mapping study',  'systematic mapping', \\ 
'emerging trends', 'comprehensive survey', 'contemporary survey',\\
'systematization of knowledge', 'systematisation of knowledge',\\ 
'tutorial', 'review of', 'research directions', 'a review',\\
'case studies', 'overview of', 'summary of'

\end{tabular}
\end{table}

\clearpage
\subsection{NCI Analysis by Paper Type, 2000–2024}
\label{app:nci_2000_2024}

\subsubsection*{Summary Descriptive Statistics}
Median values with 95\% confidence intervals.

\begin{table}[h!]
\centering
\caption{Descriptive statistics for NCI by paper type (median with 95\% confidence intervals).}
\begin{tabular}{lcc}
\toprule
 & n & Median [95\% CI] \\
\midrule
Regular & 16,056,957 & 0.437 [0.437, 0.438] \\
NR & 393,034 & 0.663 [0.663, 0.667] \\
SR & 138,260 & 0.903 [0.898, 0.911] \\
MA & 111,395 & 0.883 [0.872, 0.895] \\
\bottomrule
\end{tabular}
\end{table}

\subsubsection*{Kruskal--Wallis Test Summary}
\begin{table}[h!]
\centering
\caption{Results of the Kruskal--Wallis $H$ test comparing NCI across paper types.}
\begin{tabular}{lll}
\toprule
\textbf{Statistic} & \textbf{Value} & \textbf{Interpretation} \\
\midrule
$H$ & 99688.184 & --- \\
$p$ & 0.000 & Significant \\
$\varepsilon^2$ & 0.006 & Small effect \\
\bottomrule
\end{tabular}
\end{table}

\subsubsection*{Post-hoc Dunn Test Summary}
Pairwise comparisons (Bonferroni-adjusted $p$-values) following the Kruskal--Wallis test are reported below. Values $p < 0.01$ indicate significant pairwise differences.

\begin{table}[h!]
\centering
\caption{Dunn’s post-hoc pairwise comparison matrix for NCI.}
\begin{tabular}{lcccc}
\toprule
 & Regular & NR & SR & MA \\
\midrule
Regular & 1.000 & 0.000 & 0.000 & 0.000 \\
NR & 0.000 & 1.000 & 0.000 & 0.000 \\
SR & 0.000 & 0.000 & 1.000 & 1.000 \\
MA & 0.000 & 0.000 & 1.000 & 1.000 \\
\bottomrule
\end{tabular}
\end{table}

\subsubsection*{Pairwise Effect Sizes (Cliff’s $\delta$)}
Pairwise Cliff’s delta ($\delta$) values quantify effect size magnitudes. Interpretation: neg ($<$0.147), small ($<$0.33), medium ($<$0.474), large ($\ge$0.474).

\begin{table}[h!]
\centering
\caption{Pairwise Cliff’s delta ($\delta$) values and qualitative interpretations for NCI.}
\begin{tabular}{lcccc}
\toprule
Group 1 & Group 2 & $\delta$ & Effect size & Adj. $p$ \\
\midrule
Regular & NR & -0.190000 & small & 0.000 \\
Regular & SR & -0.290000 & small & 0.000 \\
Regular & MA & -0.290000 & small & 0.000 \\
NR & SR & -0.090000 & neg & 0.000 \\
NR & MA & -0.090000 & neg & 0.000 \\
SR & MA & 0.000000 & neg & 1.000 \\
\bottomrule
\end{tabular}
\end{table}

\subsection{NCI Analysis by Paper Type, 2000–2009}
\label{app:nci_2000_2009}

\subsubsection*{Summary Descriptive Statistics}
Median values with 95\% confidence intervals.

\begin{table}[h!]
\centering
\caption{Descriptive statistics for NCI by paper type (median with 95\% confidence intervals).}
\begin{tabular}{lcc}
\toprule
 & n & Median [95\% CI] \\
\midrule
Regular & 3,762,169 & 0.417 [0.416, 0.418] \\
NR & 83,542 & 0.599 [0.595, 0.611] \\
SR & 11,176 & 1.038 [1.010, 1.065] \\
MA & 8,894 & 1.401 [1.352, 1.455] \\
\bottomrule
\end{tabular}
\end{table}

\subsubsection*{Kruskal--Wallis Test Summary}
\begin{table}[h!]
\centering
\caption{Results of the Kruskal--Wallis $H$ test comparing NCI across paper types.}
\begin{tabular}{lll}
\toprule
\textbf{Statistic} & \textbf{Value} & \textbf{Interpretation} \\
\midrule
$H$ & 15496.917 & --- \\
$p$ & 0.000 & Significant \\
$\varepsilon^2$ & 0.004 & Small effect \\
\bottomrule
\end{tabular}
\end{table}

\subsubsection*{Post-hoc Dunn Test Summary}
Pairwise comparisons (Bonferroni-adjusted $p$-values) following the Kruskal--Wallis test are reported below. Values $p < 0.01$ indicate significant pairwise differences.

\begin{table}[h!]
\centering
\caption{Dunn’s post-hoc pairwise comparison matrix for NCI.}
\begin{tabular}{lcccc}
\toprule
 & Regular & NR & SR & MA \\
\midrule
Regular & 1.000 & 0.000 & 0.000 & 0.000 \\
NR & 0.000 & 1.000 & 0.000 & 0.000 \\
SR & 0.000 & 0.000 & 1.000 & 0.000 \\
MA & 0.000 & 0.000 & 0.000 & 1.000 \\
\bottomrule
\end{tabular}
\end{table}

\subsubsection*{Pairwise Effect Sizes (Cliff’s $\delta$)}
Pairwise Cliff’s delta ($\delta$) values quantify effect size magnitudes. Interpretation: neg ($<$0.147), small ($<$0.33), medium ($<$0.474), large ($\ge$0.474).

\begin{table}[h!]
\centering
\caption{Pairwise Cliff’s delta ($\delta$) values and qualitative interpretations for NCI.}
\begin{tabular}{lcccc}
\toprule
Group 1 & Group 2 & $\delta$ & Effect size & Adj. $p$ \\
\midrule
Regular & NR & -0.160000 & small & 0.000 \\
Regular & SR & -0.340000 & medium & 0.000 \\
Regular & MA & -0.440000 & medium & 0.000 \\
NR & SR & -0.170000 & small & 0.000 \\
NR & MA & -0.280000 & small & 0.000 \\
SR & MA & -0.130000 & neg & 0.000 \\
\bottomrule
\end{tabular}
\end{table}

\subsection{NCI Analysis by Paper Type, 2015–2024}
\label{app:nci_2015_2024}

\subsubsection*{Summary Descriptive Statistics}
Median values with 95\% confidence intervals.

\begin{table}[h!]
\centering
\caption{Descriptive statistics for NCI by paper type (median with 95\% confidence intervals).}
\begin{tabular}{lcc}
\toprule
 & n & Median [95\% CI] \\
\midrule
Regular & 9,106,489 & 0.449 [0.449, 0.449] \\
NR & 234,054 & 0.693 [0.690, 0.693] \\
SR & 108,638 & 0.858 [0.836, 0.873] \\
MA & 85,793 & 0.808 [0.808, 0.820] \\
\bottomrule
\end{tabular}
\end{table}

\subsubsection*{Kruskal--Wallis Test Summary}
\begin{table}[h!]
\centering
\caption{Results of the Kruskal--Wallis $H$ test comparing NCI across paper types.}
\begin{tabular}{lll}
\toprule
\textbf{Statistic} & \textbf{Value} & \textbf{Interpretation} \\
\midrule
$H$ & 64301.835 & --- \\
$p$ & 0.000 & Significant \\
$\varepsilon^2$ & 0.007 & Small effect \\
\bottomrule
\end{tabular}
\end{table}

\subsubsection*{Post-hoc Dunn Test Summary}
Pairwise comparisons (Bonferroni-adjusted $p$-values) following the Kruskal--Wallis test are reported below. Values $p < 0.01$ indicate significant pairwise differences.

\begin{table}[h!]
\centering
\caption{Dunn’s post-hoc pairwise comparison matrix for NCI.}
\begin{tabular}{lcccc}
\toprule
 & Regular & NR & SR & MA \\
\midrule
Regular & 1.000 & 0.000 & 0.000 & 0.000 \\
NR & 0.000 & 1.000 & 0.000 & 0.000 \\
SR & 0.000 & 0.000 & 1.000 & 0.000 \\
MA & 0.000 & 0.000 & 0.000 & 1.000 \\
\bottomrule
\end{tabular}
\end{table}

\subsubsection*{Pairwise Effect Sizes (Cliff’s $\delta$)}
Pairwise Cliff’s delta ($\delta$) values quantify effect size magnitudes. Interpretation: neg ($<$0.147), small ($<$0.33), medium ($<$0.474), large ($\ge$0.474).

\begin{table}[h!]
\centering
\caption{Pairwise Cliff’s delta ($\delta$) values and qualitative interpretations for NCI.}
\begin{tabular}{lcccc}
\toprule
Group 1 & Group 2 & $\delta$ & Effect size & Adj. $p$ \\
\midrule
Regular & NR & -0.200000 & small & 0.000 \\
Regular & SR & -0.270000 & small & 0.000 \\
Regular & MA & -0.250000 & small & 0.000 \\
NR & SR & -0.060000 & neg & 0.000 \\
NR & MA & -0.040000 & neg & 0.000 \\
SR & MA & 0.030000 & neg & 0.000 \\
\bottomrule
\end{tabular}
\end{table}

\subsection{NCI Between Periods}
\label{app:nci_beween_periods}
We assessed differences in nci between Early Period (2000--2009) and Late Period (2015--2024) for each paper type using Mann--Whitney U tests. Cliff's delta (\(\delta\)) quantifies the direction and magnitude of change, while adjusted p-values (via Bonferroni) control for multiple testing ($\alpha$ = 0.01). Medians are shown with 95\% confidence intervals.\subsubsection{Summary of statistical results}
\begin{table}[h!]
\centering
\caption{Per-type comparison of NCI between Early Period and Late Period using Mann--Whitney U tests (Bonferroni-adjusted $p$-values). Medians are shown with 95\% confidence intervals.}
\label{tab:nci-periods}
\begin{tabular}{llllll}
\toprule
Type & med Early & med Late & U & p & delta \\
\midrule
Reg & 0.417 [0.416, 0.418] & 0.449 [0.449, 0.449] & 17237001763666.000 & 0.000 & 0.006 \\
NR & 0.599 [0.595, 0.611] & 0.693 [0.690, 0.693] & 9477837969.000 & 0.000 & -0.031 \\
SR & 1.038 [1.010, 1.065] & 0.858 [0.836, 0.873] & 657328240.000 & 0.000 & 0.083 \\
MR & 1.401 [1.352, 1.455] & 0.808 [0.808, 0.820] & 469921687.000 & 0.000 & 0.232 \\
\bottomrule
\end{tabular}
\end{table}

\clearpage

\subsection{AI-Score Analysis by Field of Study}
\label{app:ai_score_by_fos}

\begin{table}[h!]
\centering
\caption{Summary of AI-score results by Field of Study (median with 95\% confidence intervals; top 23 fields).}
\label{tab:fostable}
\begin{tabular}{lllrll}
\toprule
FoS & count & Median [95\% CI] & n(higher) & Mean $\delta$ & Median $\delta$ \\
\midrule
Business & 911 & 2.252 [2.121, 2.357] & 21 & 0.610 & 0.640 \\
Art & 196 & 2.018 [1.797, 2.186] & 21 & 0.580 & 0.600 \\
Law & 438 & 1.478 [1.365, 1.640] & 14 & 0.410 & 0.440 \\
Linguistics & 678 & 1.428 [1.361, 1.540] & 14 & 0.410 & 0.430 \\
Computer Sci & 41,998 & 1.415 [1.407, 1.424] & 14 & 0.430 & 0.450 \\
Education & 2,354 & 1.402 [1.362, 1.459] & 14 & 0.390 & 0.410 \\
Political Sci & 501 & 1.321 [1.249, 1.430] & 13 & 0.350 & 0.380 \\
History & 238 & 1.278 [1.143, 1.481] & 11 & 0.340 & 0.340 \\
Geography & 164 & 1.259 [1.118, 1.431] & 9 & 0.400 & 0.390 \\
Materials Sci & 9,212 & 1.104 [1.082, 1.124] & 11 & 0.230 & 0.240 \\
Philosophy & 186 & 1.091 [0.939, 1.317] & 7 & 0.270 & 0.250 \\
Economics & 935 & 1.070 [1.019, 1.119] & 9 & 0.220 & 0.220 \\
Chemistry & 9,771 & 1.061 [1.053, 1.078] & 9 & 0.230 & 0.230 \\
Engineering & 22,585 & 1.042 [1.030, 1.051] & 9 & 0.220 & 0.220 \\
Env. Sci & 17,897 & 0.958 [0.941, 0.972] & 7 & 0.170 & 0.160 \\
Ag. \& Food Sci & 4,364 & 0.891 [0.869, 0.907] & 4 & 0.160 & 0.150 \\
Biology & 10,511 & 0.866 [0.855, 0.880] & 4 & 0.140 & 0.130 \\
Sociology & 2,790 & 0.858 [0.829, 0.887] & 3 & 0.130 & 0.130 \\
Geology & 262 & 0.836 [0.733, 0.963] & 1 & 0.160 & 0.160 \\
Psychology & 6,512 & 0.805 [0.792, 0.821] & 2 & 0.130 & 0.130 \\
Medicine & 110,819 & 0.795 [0.791, 0.805] & 2 & 0.090 & 0.090 \\
Mathematics & 2,108 & 0.726 [0.694, 0.753] & 0 &  &  \\
Physics & 10,657 & 0.670 [0.660, 0.677] & 0 &  &  \\
\bottomrule
\end{tabular}
\end{table}

\clearpage

\subsection{Mann--Whitney Analysis by Field of Study}
\label{app:ai_score_by_fos_reg_vs_rev_mann_whitney}
Each field was compared using a Mann--Whitney U test, reporting medians with 95\% Hettmansperger-Sheather confidence intervals. P-values were Bonferroni-adjusted (\(\alpha=0.05\)), and effect sizes were expressed as Cliff's \(\delta\).\\[6pt]
\textbf{Pooled overall test:} $U=2095156969.000$, $p=0.044$, $\delta=0.008$ (negligible effect).\\
\textbf{Stratified test:} $Z=21.309$, $p=0.000$ (mean $\delta=0.130$; median $\delta=0.162$) based on 13 fields.
\begin{table}[htbp]
\centering
\caption{Per-field Mann--Whitney U test results comparing AI-scores between review and regular papers (medians with 95\% CIs), sorted by overall median AI-score.}
\begin{tabular}{lrrllrc}
\toprule
FoS & n Reg & n Rev & med Reg & med Rev & $\delta$ & Adj. $p$ \\
\midrule
Business & 808 & 103 & 2.27 [2.13, 2.40] & 2.06 [1.80, 2.45] & -0.03 & 1.00 \\
Computer Sci & 40,476 & 1,522 & 1.41 [1.40, 1.42] & 1.79 [1.72, 1.85] & 0.20 & 0.00 \\
Education & 2,133 & 221 & 1.37 [1.33, 1.44] & 1.63 [1.47, 1.72] & 0.09 & 0.21 \\
Materials Sci & 8,815 & 397 & 1.09 [1.07, 1.12] & 1.36 [1.26, 1.50] & 0.19 & 0.00 \\
Chemistry & 9,475 & 296 & 1.05 [1.04, 1.07] & 1.37 [1.25, 1.52] & 0.19 & 0.00 \\
Engineering & 21,761 & 824 & 1.03 [1.02, 1.04] & 1.48 [1.40, 1.59] & 0.27 & 0.00 \\
Env. Sci & 16,947 & 950 & 0.95 [0.93, 0.96] & 1.13 [1.07, 1.21] & 0.13 & 0.00 \\
Ag. \& Food Sci & 3,993 & 371 & 0.87 [0.85, 0.90] & 1.17 [1.02, 1.31] & 0.17 & 0.00 \\
Biology & 10,222 & 289 & 0.87 [0.85, 0.88] & 1.05 [0.95, 1.15] & 0.16 & 0.00 \\
Sociology & 2,587 & 203 & 0.86 [0.83, 0.90] & 0.84 [0.75, 0.91] & -0.03 & 1.00 \\
Psychology & 5,930 & 582 & 0.79 [0.78, 0.82] & 0.94 [0.86, 0.98] & 0.09 & 0.00 \\
Medicine & 98,820 & 11,999 & 0.79 [0.79, 0.80] & 0.84 [0.82, 0.85] & 0.04 & 0.00 \\
Physics & 10,527 & 130 & 0.67 [0.66, 0.68] & 1.01 [0.85, 1.20] & 0.21 & 0.00 \\
\bottomrule
\end{tabular}
\end{table}

\clearpage

\subsection{Excess AI Score Analysis by Paper Type, 2024–2024}
\label{app:excess ai score_2024_2024}

\subsubsection*{Summary Descriptive Statistics}
Median values with 95\% confidence intervals.

\begin{table}[h!]
\centering
\caption{Descriptive statistics for Excess AI Score by paper type (median with 95\% confidence intervals).}
\begin{tabular}{lcc}
\toprule
 & n & Median [95\% CI] \\
\midrule
Regular & 232,494 & 0.974 [0.972, 0.979] \\
NR & 12,043 & 1.073 [1.053, 1.098] \\
SR & 3,950 & 0.926 [0.896, 0.967] \\
MA & 1,894 & 0.676 [0.663, 0.698] \\
\bottomrule
\end{tabular}
\end{table}

\subsubsection*{Kruskal--Wallis Test Summary}
\begin{table}[h!]
\centering
\caption{Results of the Kruskal--Wallis $H$ test comparing Excess AI Score across paper types.}
\begin{tabular}{lll}
\toprule
\textbf{Statistic} & \textbf{Value} & \textbf{Interpretation} \\
\midrule
$H$ & 661.636 & --- \\
$p$ & 0.000 & Significant \\
$\varepsilon^2$ & 0.003 & Small effect \\
\bottomrule
\end{tabular}
\end{table}

\subsubsection*{Post-hoc Dunn Test Summary}
Pairwise comparisons (Bonferroni-adjusted $p$-values) following the Kruskal--Wallis test are reported below. Values $p < 0.01$ indicate significant pairwise differences.

\begin{table}[h!]
\centering
\caption{Dunn’s post-hoc pairwise comparison matrix for Excess AI Score.}
\begin{tabular}{lcccc}
\toprule
 & Regular & NR & SR & MA \\
\midrule
Regular & 1.000 & 0.000 & 0.001 & 0.000 \\
NR & 0.000 & 1.000 & 0.000 & 0.000 \\
SR & 0.001 & 0.000 & 1.000 & 0.000 \\
MA & 0.000 & 0.000 & 0.000 & 1.000 \\
\bottomrule
\end{tabular}
\end{table}

\subsubsection*{Pairwise Effect Sizes (Cliff’s $\delta$)}
Pairwise Cliff’s delta ($\delta$) values quantify effect size magnitudes. Interpretation: neg ($<$0.147), small ($<$0.33), medium ($<$0.474), large ($\ge$0.474).

\begin{table}[h!]
\centering
\caption{Pairwise Cliff’s delta ($\delta$) values and qualitative interpretations for Excess AI Score.}
\begin{tabular}{lcccc}
\toprule
Group 1 & Group 2 & $\delta$ & Effect size & Adj. $p$ \\
\midrule
Regular & NR & -0.070000 & neg & 0.000 \\
Regular & SR & 0.040000 & neg & 0.001 \\
Regular & MA & 0.290000 & small & 0.000 \\
NR & SR & 0.100000 & neg & 0.000 \\
NR & MA & 0.330000 & medium & 0.000 \\
SR & MA & 0.240000 & small & 0.000 \\
\bottomrule
\end{tabular}
\end{table}

\end{document}